\documentstyle[epsfig,aps,prb,twocolumn,floats,eqsecnum]{revtex}

\begin{document}

\newcommand{\gsim}{
\,\raisebox{0.35ex}{$>$}
\hspace{-1.7ex}\raisebox{-0.65ex}{$\sim$}\,
}

\newcommand{\lsim}{
\,\raisebox{0.35ex}{$<$}
\hspace{-1.7ex}\raisebox{-0.65ex}{$\sim$}\,
}

\newcommand{\const}{ {\rm const} }
\newcommand{\arctanh}{ {\rm arctanh} }

\bibliographystyle{prsty}

\title{ \begin{flushleft}
{\small \em published in}\\
{\small
PHYSICAL REVIEW LETTERS
\hfill
VOLUME 77
NUMBER 25
\hfill
16 DECEMBER 1996
}\\
\end{flushleft}
Kondo Screening and Magnetic Ordering in Frustrated UNi$_4$B
}

\author{
Claudine Lacroix$^a$ \cite{e-lac}, Benjamin~Canals$^a$ \cite{e-can}, and M. D. N\'u\~nez-Regueiro$^b$ \cite{e-nun} 
}

\address{
\smallskip
$^a$Laboratoire de Magn\'etisme Louis N\'eel, CNRS, Bo\^ite Postale 166, 38042 Grenoble Cedex 9, France\\
\smallskip
$^b$European Synchrotron Radiation Facility, Bo\^ite Postale 220, 38043 Grenoble Cedex, France\\
\smallskip
{\rm (Received 20 June 1996) }
\bigskip\\
\parbox{14.2cm}
{\rm
UNi$_4$B exhibits unusual properties and, in particular, a unique antiferromagnetic arrangement involving only $2/3$ of the U sites. 
Based on the low temperature behavior of this compound, we propose that the remaining $1/3$ U sites are nonmagnetic due to the Kondo effect. 
We derive a model in which the coexistence of magnetic and nonmagnetic U sites is the consequence of the competition between frustration of the crystallographic structure and instability of the $5f$ moments.
\smallskip
\begin{flushleft}
PACS number(s): 75.10.-b, 75.30.Mb
\end{flushleft}
} 
} 
\maketitle

In a series of recent papers \cite{men941,men951,men952,men93} the interesting magnetic behavior of UNi$_4$B was discussed. 
This intermetallic compound crystallizes in a CeCo$_4$B-type structure, space group $P6/mmm$, in which UNi and UB planes alternate with Ni planes in between, see Fig. 1(a). 
Only the U atoms have a magnetic moment, and they display a hexagonal arrangement in the basal plane. 
The distance between nearest-neighbor (nn) U atoms in the basal plane ($a =
4.95 $ \AA) is larger than in the perpendicular direction ($c/2 = 3.48 $ \AA), forming a triangular lattice of ferromagnetic (F) chains. 
The puzzling properties of this compound are certainly reflecting the geometrical frustration of this triangular lattice with antiferromagnetic (AF) interactions.

Below $T_N = 20$ K neutron diffraction experiments \cite{men941,men951} indicate that only $2/3$ of the U moments order antiferromagnetically in a complex structure, the magnetic unit cell involving nine U atoms, as can be seen in Fig. 1(b). 
The six ordered magnetic U moments are perpendicular to the {\it c}-axis forming a 120$^o$ angle between next-nearest neighbors (nnn). 
The application of a magnetic field along the {\it c}-axis direction or parallel to the basal plane shows the strong anisotropy of the system. 
In the basal plane, two or three (more cannot be excluded) jumps for the magnetization are observed, depending on the field direction \cite{men951}.

This structure has been interpreted \cite{men951} assuming two independent spin systems, one which orders while the other remains paramagnetic down to low temperatures. 
It was proposed that $1/3$ of the U atoms forms chains within the ordered spin matrix that keep their one-dimensional character because the local field vanishes on those sites. 
However the expected ordering of these ``paramagnetic'' sites when a small field is applied is not experimentally observed, suggesting another explanation for these $1/3$ U atoms.

\begin{figure}[t]
\unitlength1cm
\begin{picture}(11,12)
\centerline{\epsfig{file=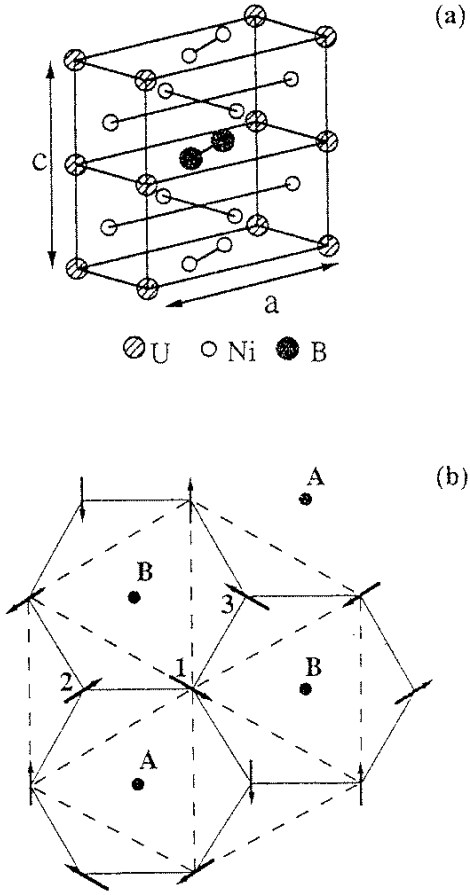,angle=0,width=6.5cm}}
\end{picture}
\\
\caption{
(a) Crystallographic CeCo$_4$B-type subcell of UNi$_4$B. (b) Zero-field magnetic structure of UNi$_4$B projected on the basal plane. The arrows indicate the magnetic U atoms, while the solid circles represent the Kondo screened U sites. These layers are stacked feromagnetically along the {\it c}-axis. The dashed lines relates sites of one of the three independent lattices determined when only nnn $J_2$ interactions are considered. Lattice distortions differentiate $A$ and $B$ nonmagnetic sites, reducing (increasing) the distance between 1 and 2 (1 and 3) magnetic U atoms.
}
\end{figure}

We would like to point out that, in itinerant systems close to a magnetic-nonmagnetic (M-NM) instability, frustration can be avoided or diminished by partial vanishing of the magnetic moments.
Considering the low temperature behavior of UNi$_4$B we propose in this Letter that its ordered structure occurs because $1/3$ of the U moments are canceled by Kondo compensation. 
We derive a model which takes into account this Kondo effect and that, when applied to the hexagonal structure of UNi$_4$B, allows the interpretation of its properties.

In fact several properties are very similar to those observed in some Laves-phase RMn$_2$ compounds \cite{yos92}. 
In DyMn$_2$, ThMn$_2$, and also in TbMn$_2$ under applied field, ordered ``mixed'' phases with the coexistence of magnetic and nonmagnetic Mn sites have been found. 
This is the peculiarity of these intermetallic systems with respect to other frustrated systems, largely discussed in literature \cite{cha95}, in which all sites are always magnetic. 
This behavior has been explained \cite{bal91} by the interplay between frustration and the instability of the {\it 3d}-Mn magnetic moment. 
But this is a more general phenomenon: The M-NM instability can also result from the Kondo effect or in compounds where the lowest crystal field level is a singlet.

As remarked in Refs. [\onlinecite{men952,men942,men}] the existence of a Kondo effect in UNi$_4$B can be inferred from its low temperature properties: 
(i) a continuous rise of the basal-plane resistivity upon lowering the temperature; 
(ii) an enhancement of the linear electronic specific-heat coefficient $\gamma$ in the AF state, $\gamma = 250$ mJ/mole K$^2$, which is a large value if it is attributed to the nonmagnetic U sites; 
(iii) a large negative Curie-Weiss temperature $\theta = -65$ K that can only be explained by a Kondo effect, since the dominant interactions, which are those within the chains, are ferromagnetic; 
(iv) below $T_N$ there is an increase of the susceptibility which is suppressed by increasing applied magnetic fields \cite{men93}.

Since the strongest interactions are those along the {\it c}-axis, we first consider the uranium chains: We describe them by Kondo lattice chains close to the ferromagnetic-Kondo instability. 
In mean-field approximation (MFA) \cite{lac79}, the energy of a ferromagnetic U chain is $E_F = -J^2 / 32 W$, while the energy of the Kondo state is given by $E_K$~=~$-nW \exp(2W/J)$. 
We call $J$ the local exchange energy between the $5f$ U moment and the conduction electrons, $W$ the half bandwidth of the band, and $n$ the concentration of conduction electrons. 
We consider the case $n > |J/4W|$ in order to avoid the problem of small concentrations: Close to the bottom of the band the results are very sensitive to its shape. 
The results of Ref. [\onlinecite{lac79}] have been obtained for a constant density of states, while for the strictly one-dimensional case, the divergence at the band edge leads to a different phase diagram \cite{rem}. 
As the ordering in the real compound is, in fact, three dimensional, the MFA is probably more appropriate for describing UNi$_4$B (i.e., the one-dimensional character does not seem to be crucial in the ordered phase). 
In any case, for our purpose the only important feature is the proximity to the instability, i.e., $|E_F - E_K|$ small.

Next, the effective hopping $t_{ij}$ between nn (or nnn) chains is considered. 
We calculate the energy interaction between U chains in second order perturbation in $t_{ij}$. Depending on the configuration of the chains, the four interactions energies for a small value of $J/W$ are as follows:

\begin{eqnarray}
E_{KK} & = & 2 E_K - \frac{ t_{ij}^{2} }{W} (1 - \frac{ J^2 {\lambda}^2 }{ nW^2 } (1+n^2)),	\nonumber \\
E_{KF} & = &  E_K + E_F - \frac{ t_{ij}^{2} }{ W } (1 - 4 \frac{ J {\lambda}^2 }{ nW } + 8 \lambda \ln n),  \nonumber \\
E_{\uparrow \uparrow} & = &  2 E_F, \nonumber \\
E_{\uparrow \downarrow} & = &  2 E_F - \frac{ t_{ij}^{2} }{ W },
\end{eqnarray}
where $\lambda^2 = n(W/J)^2 e^{2W/J}$ is, in that case, a small parameter [only the leading corrections in $\lambda^2$ are considered in Eq. (1)]. 
These energies correspond to two Kondo, one Kondo and one F, and two F chains with equal or opposite spin orientation, respectively. 
This yields to the following effective Hamiltonian:

\begin{eqnarray}
H = & & \sum_{i} {({\Delta}_i + D {\cos}^2 3 \theta_i) |{\mu}_{i}^{2}|} - \sum_{i \neq j} {J_{ij} {\vec{\mu}}_{i}.\vec{{\mu}}_{j}} \nonumber \\
    & + & \sum_{i \neq j} {V_{ij} ({\mu}_0 - |{\mu}_i|) ({\mu}_0 - |{\mu}_j|))},
\end{eqnarray}
where $\mu_i$ is the effective U magnetic moment of the $i$ chain. 
If there is a Kondo effect, $\mu_i = 0$ because of the compensation by the conduction band. 
This procedure allows us to consider both magnetic and nonmagnetic sites simultaneously, in contrast with the previous analysis \cite{men951}. 
At high temperature the U magnetic moment is estimated $\mu_i = 2.9 \mu_B /U$ atom but below $T_N$, in the ordered phase, the effective moment of the magnetic sites is $\mu_0 = 1.2 \mu_B /U$ atom \cite{men951}. 
The last term in Eq. (0.2) describes the repulsion between NM chains, and it vanishes when one of the chains is F.

The parameters of Eq. (0.2) are related to the energies of Eq. (0.1) in the following way [neglecting corrections in $\lambda^2$, which can eventually be easily calculated from Eq. (0.1)]:

\begin{eqnarray}
J_{ij} & = & \frac{1}{2} (E_{\uparrow \downarrow} - E{\uparrow \uparrow}) \approx - \frac{t_{ij}^{2}}{2W},	\nonumber \\
{\Delta}_i & = & \frac{1}{2} (E_{\uparrow \downarrow} + E_{\uparrow \uparrow}) - E_{KF} \approx E_F - E_K + \sum_{j} {\frac{t_{ij}^{2}}{2W}}, \\
V_{ij} & = & E_{KK} - 2 E_{KF} + \frac{1}{2} (E_{\uparrow \downarrow} + E_{\uparrow \uparrow}) \approx \frac{t_{ij}^{2}}{2W}. \nonumber
\end{eqnarray}

The $D$ ($D > 0$) term describes the crystalline anisotropy in the hexagonal lattice, $\theta_i$ is the angle of spins with the $x$ axis. 
As noticed in Ref. [\onlinecite{men951}],the steps in the magnetization curves are an experimental indication of the relevance of the anisotropy in this compound. 
Furthermore, it has been shown theoretically \cite{reg94} that the anisotropy is a crucial parameter for the stabilization of ``mixed phases'': If $D = 0$ an incommensurate helimagnetic structure will always have lower energy.

The energy necessary to create a magnetic chain, $\Delta_i = \Delta$, will be taken as positive. Since it is related to the Kondo effect, $\Delta$ is a temperature-dependent parameter.

The $J_{ij}$ terms describe the nn ($J_1$) and nnn ($J_2$) exchange interactions, in this calculation they are similar to AF superexchange interactions. 
Longer-range interactions are assumed to be less important, and will thus be neglected.

A repulsion $V_{ij}$ between nonmagnetic chains is also obtained in the calculation. 
Since it is of the same order as the exchange energy, it must be considered. 
Furthermore, in the calculations, a nn $V_{ij} = V$ appears to be necessary in order to stabilize the experimental magnetic ordering shown in Fig. 1(b).

In Ref. [\onlinecite{men951}] the magnetic structure was explained by considering only nnn interactions $J_2$ and the anisotropy $D$. 
However, in that case, the two sublattices formed by the nnn magnetic chains can rotate independently of each other without energy cost. 
The degeneracy can be completely lifted by the observed lattice distortions \cite{men951,dro95}, approaching the magnetic U atoms to the nonmagnetic A sites, see Fig. 1(b). 
If $J_1$ depends on the U-U distance, this leads to different $J_1$ interactions.

In Fig. 2 the region where the observed configuration can take place is shown in the phase diagram $J_1 / \Delta$ vs $J_2 / \Delta$ for a given value of $V / \Delta$ and for infinite $D$. 
This phase diagram has been obtained by comparing the energies of all possible ordered phases with a magnetic unit cell not larger than nine sites. 
The mixed structure is stable for AF $J_2$, large enough to avoid the NM phase. For $J_2 / \Delta < -1/3$, all sites become magnetic because the energy gained by Kondo screening cannot compensate the exchange energy. The same occurs with incrasing $|J_1|$.
In contrast, for ferromagnetic $J_2$, the usual magnetic ordered phases of the triangular lattice are obtained (either F for $J_1 > 0$ or ``120$^o$'' for $J_1 < 0$), except for very small $J_1 / \Delta$ and $J_2 / \Delta$, where all moments are screened by the Kondo effect. 
However, without distortions, the experimental phase is degenerate with six other ordered structures, all having the same number of NM chains. 
If $J_1$ decreases with distance $J_{1}^{1-2} > J_{1}^{1-3}$, and the observed configuration is stabilized.
\begin{figure}[t]
\unitlength1cm
\begin{picture}(15,8)
\centerline{\epsfig{file=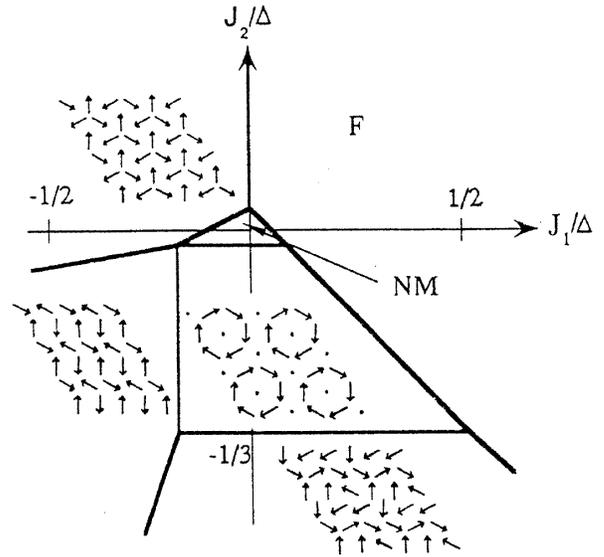,angle=0,width=8cm}}
\end{picture}
\\
\caption{
Phase diagram of Eq. (0.2) for $V/\Delta = 0.1$ and $D = \infty$. NM, nonmagnetic phase; F, ferromagnetic phase.
}
\end{figure}

When a magnetic field was applied in the basal plane, two or three steps were observed \cite{men951}, depending on the field direction. 
In the last case, saturation was not attained up to 50 T. 
This behavior is well reproduced by Eq. (0.2) with $J_1 \approx 0.1 K$, $J_2 \approx - 1.3 K$, $\Delta \approx 9 K$. 
This value of $\Delta$ is in good agreement with estimations of the Kondo temperature deduced from specific heat and thermal expansion measurements \cite{men942,men}. 
In low fields, the steps are due to the reorientation of the spins, but, in higher fields, transition from Kondo to magnetic chains can occur. 
In the experimental case, the jumps are small and not sharp, see Fig. 2 in Ref. [\onlinecite{men951}]. 
These may have different origins: 
(i) the anisotropy is finite, 
(ii) a magnetic moment is induced on the Kondo chains by the field, 
(iii) the numerical calculations show that a finite $J_1$ interaction drives several intermediate phases that smooth the transitions. 
On the other hand, the nonsaturation of the magnetization in the hard-basal direction indicates that the anisotropy $D$ is certainly larger than estimated in Ref. [\onlinecite{men951}], where the Kondo screening was not considered. 
It would be interesting to compare saturation fields in both in-plane directions to have a more accurate determination of this anisotropy. 
Both effects (i) and (ii) contribute to the low-field susceptibility and must be simultaneously taken into account when evaluating the parameters. 
Neutron experiments under field would allow one to understand the intermediate magnetic phases, which depend on the relative values of $D$, $\Delta$, $J_1$ and $J_2$, and to test the validity of our model. 
In a high enough applied field the disappearance of the Kondo effect should be observed.

In conclusion, the unusual behavior of UNi$_4$B exhibits many similarities with the RMn$_2$ compounds. 
It is another example of the effect of frustration in itinerant systems, in which, although all U sites are, in principle, equivalent, some of them are canceled to stabilize a mixed ordered structure. 
It can be described by the general Hamiltonian, Eq. (0.2); in this case, the partial vanishing of the magnetic sites is driven by the Kondo effect. 
We have shown how this Hamiltonian can be derived from a microscopic model; in practice, the effective parameters of Eq. (0.2) are sufficient. 
While the effect of frustration has been largely discussed in insulating systems, many of its consequences in the metallic case have not been addressed. 
It would be worthwhile to find other examples of these systems in order to have an overall picture of this new phenomena.


\end{document}